\documentclass{article}

\usepackage{microtype}
\usepackage{graphicx}
\usepackage{subfigure}
\usepackage{booktabs} 
\usepackage{physics}
\usepackage{newtxtext, newtxmath}
\usepackage{amsmath}
\usepackage[T1]{fontenc}
\usepackage{ae,aecompl}
\usepackage{bm}
\usepackage{hyperref}

\usepackage[accepted]{icml2021}

\DeclareMathOperator*{\argmin}{arg\,min}

\icmltitlerunning{Graph-GAN for Galaxies}

\begin{document}

\twocolumn[
\icmltitle{Galaxies on graph neural networks: towards robust synthetic galaxy catalogs with deep generative models.}



\icmlsetsymbol{equal}{*}

\begin{icmlauthorlist}
\icmlauthor{Yesukhei Jagvaral}{cmu,cmu-ai}
\icmlauthor{Fran\c{c}ois Lanusse}{cnrs}
\icmlauthor{Sukhdeep Singh}{cmu,cmu-ai}
\icmlauthor{Rachel Mandelbaum}{cmu,cmu-ai}
\icmlauthor{Siamak Ravanbakhsh}{mila}
\icmlauthor{Duncan Campbell}{cmu,epi}
 
\end{icmlauthorlist}

\icmlaffiliation{cmu}{McWilliams Center for Cosmology, Department of Physics, Carnegie Mellon University, Pittsburgh, PA 15213, USA}
\icmlaffiliation{cmu-ai}{NSF AI Planning Institute for Data-Driven Discovery in Physics, Carnegie Mellon University, Pittsburgh, PA 15213, USA}
\icmlaffiliation{cnrs}{AIM, CEA, CNRS, Universit\'e Paris-Saclay, Universit\'e Paris Diderot, Sorbonne Paris Cit\'e, F-91191 Gif-sur-Yvette, France}
\icmlaffiliation{mila}{Mila, Quebec AI Institute}
\icmlaffiliation{epi}{Epistemix Inc., Pittsburgh}

\icmlcorrespondingauthor{Yesukhei Jagvaral}{yjagvara@andrew.cmu.edu}
 
\icmlkeywords{Machine Learning, ICML}

\vskip 0.7in
]



\printAffiliationsAndNotice{}  
\begin{abstract}
The future astronomical imaging surveys are set to provide precise constraints on cosmological parameters, such as dark energy. However, production of synthetic data for  these surveys, to test and validate analysis methods, suffers from a very high computational cost. In particular, generating mock galaxy catalogs at sufficiently large volume and high resolution will soon become computationally unreachable. In this paper, we address this problem with a Deep Generative Model to create robust mock galaxy catalogs that may be used to test and develop the analysis pipelines of future weak lensing surveys.  We build our model on a custom built Graph Convolutional Networks, by placing each galaxy on a graph node and then connecting the graphs within each gravitationally bound system. We train our model on a cosmological simulation with realistic galaxy populations to capture the 2D and 3D orientations of galaxies. The samples from the  model exhibit comparable statistical properties to those in the simulations. To the best of our knowledge, this is the first instance of a generative model on graphs in an astrophysical/cosmological context.

\end{abstract}

\section{Introduction}
\label{submission}
Upcoming astronomical imaging surveys such as the Vera C.\ Rubin Observatory Legacy
Survey of Space and Time (LSST)\footnote{ \url{https://www.lsst.org/} }, Roman Space Telescope\footnote{ \url{https://roman.gsfc.nasa.gov/} } High Latitude Survey (HLS) and Euclid\footnote{ \url{https://www.euclid-ec.org/} } will aim to answer fundamental questions about the nature of dark matter and dark energy, by precisely measuring the distribution and properties of billions of galaxies. 

The analysis of these surveys will require having an access to large scale cosmological simulations for a variety of applications, ranging from validating analysis pipelines \cite{buzzard-1, buzzard-2} to constraining cosmology through Simulation-Based Inference \citep[SBI;][]{des-lfi}. 
However, as the volume and data quality of future surveys increases, cosmological simulation must cover increasingly large volumes at high resolution \citep{Vogelsberger-review}. Full hydrodynamical simulations, which can resolve the formation and evolution of individual galaxies, are extremely expensive and cannot scale to such volumes. 
This motivates the need for emulation methods capable of generating realistic mock galaxy catalogs without relying on a full simulation. One traditional solution to this problem has been to (semi-)analytically paint galaxies on N-body (gravity only) simulations. However, the assumptions behind these (semi-)analytical models may not be robust and need validation from non-parametric models \citep{somerville-dave}.
%
While machine learning would be an appealing solution to this problem, One of the main difficulties in building ML-based non-parametric models for such simulations is the fact that the data to emulate is catalog-based (i.e., a catalog of galaxy positions and properties in the simulation volume) and that each galaxy cannot be treated independently if important correlations between galaxies are to be preserved.

In this work, we propose to address this emulation problem with a conditional deep generative model, capable of modeling relevant galaxy properties and their inter-dependencies, conditioned on the underlying large-scale structure scaffolding. 
Our model combines a customized graph convolutional network architecture, to model the correlations between galaxies, with a Wasserstein Generative Adversarial Network to build a probabilistic model of galaxy properties. To the best of our knowledge, this work is the first instance of a deep generative model on graphs introduced in astrophysics.

We apply our model to the particularly challenging problem of modeling  the intrinsic alignments of galaxies. The model is able to learn and predict both 
(a) scalar features such as galaxy shapes, and more importantly, (b) correlated vector orientations in 3D and in 2D to a good quantitative agreement. 

\section{Related Work}

In the literature there is substantial work on galaxy property emulators. 
In some cases  \cite{painting-gals-bunch, modi-nn-light-dark, paco-cn} the approach has been to 
'paint' galaxy properties onto N-body (dark matter -- gravity only) simulations. However, these methods typically do not model correlations between galaxies and only predict scalar quantities, as opposed to our model which predicts both vector and scalar quantities.

While graph neural networks have been proposed in the context of cosmological simulations in previous work, it is the first time that they are used to build generative models for galaxy properties. 
\citet{cranmer-gnn} trained a graph neural network on a cosmological simulation and then extracted symbolic equations pertaining to physical laws.    \citet{  paco-gnn}, on the other hand, used graph neural networks to infer halo masses. 



\section{Directional Graph Convolutional Networks}\label{neural_model}
A key feature of our problem is that the neural architecture needs to  model direction- and distance-dependent correlations between galaxies. Instead of relying on a generic message-passing approach to build a graph neural network, we use our physical insight to build an architecture with explicit dependence on relative distance and orientation between graph nodes, as described below.
In this work, we are considering undirected and connected graphs:  $\mathcal{G} = (\mathcal{V} , \mathcal{E}, \mathbf{W})$, where $\mathcal{V}$ is the set of graphs vertices, with $\left\vert \mathcal{V} \right\vert = n$ the number of vertices,  $\mathcal{E}$ is the set of graph edges and $\mathbf{W} \in \mathbb{R}^{n \times  n}$ is the weighted adjacency matrix. 
We adopt a first order approximation to parameterize graph convolutions \citep{kipf-welling}, and define one Graph Convolutional Network layer with an activation $y_i$ for a node $i$  as:
\begin{equation}
	\forall i \in \mathcal{V}, \	   y_i = \mathbf{b} + \mathbf{W}_0 h_i + \sum_{j \in \mathcal{N}_i} w_{i, j} \mathbf{W}_1 h_j
\end{equation}
where  $\mathbf{b} $  represents a vector of bias terms; $\mathcal{N}_i$ denotes the set of immediate neighbors\footnote{Immediate neighbors or first neighbors are neighbors that are one hop away from node $i$.} of vertex $i$; $\mathbf{W}_0$ are the weights that apply a linear transform to the activation vector $h_i$ of node $i$ (i.e., self connection); $w_{i, j}$ are linear transforms on the activation vectors $h_j$ of the nodes $j$ in the neighborhood of $i$; and $\mathbf{W}_1$ are the set of weights that apply to the immediate neighbors. 
 
Following an approach proposed in \citet{Verma2017}, we implement direction-dependent graph convolution layer as
\begin{equation}
	y_i=  \mathbf{b} + \mathbf{W}_0 h_i +  \sum\limits_{m=1}^{M} \frac{1}{|\mathcal{N}_i|} \sum_{j \in \mathcal{N}_i} q_m(\mathbf{r}_i, \mathbf{r}_j) \mathbf{W}_m h_j. 
\end{equation}
Here $|\mathcal{N}_i|$ denotes the cardinality of the set $\mathcal{N}_i$, $M$ is the number of directions, and $\mathbf{r}_i$ are the 3D Cartesian coordinates of the node. 
The  $q_m(\mathbf{r}_i, \mathbf{r}_j)$ 
are normalized so that $\sum_{m=1}^{M} q_m(\mathbf{r}_i, \mathbf{r}_j) = 1$ and are defined as:
\begin{equation}
		q_m(\mathbf{r}_i, \mathbf{r}_j) \propto \exp(- \mathbf{d}_m^t \cdot (\mathbf{r}_j - \mathbf{r}_i)) \  g_\lambda( \parallel \mathbf{r}_i - \mathbf{r}_j \parallel_2^2),
\end{equation}
where the $\{ \mathbf{d}_m \}_{m \in [1,M]}$ 
are a set of directions we want to make the kernel sensitive to, and $g_\lambda$ is a parametric function of the distance  between two vertices. This can be seen as a hard-coded direction-dependent attention mechanism allowing the model to gain directional awareness by design. We further chose an exponential parametrization of the form $g_\lambda(r) = \exp( - r^2/2\lambda^2)$ for the distance-dependence, where $\lambda$ is fit automatically during training. Note that more generic functions could be used,  but this empirical parametrization was found to work well for our problem.
 

 
\section{Generative Model with Graphs}

Our goal is to learn  and sample from a conditional probability density $p(\bm{y} | \bm{x})$, where $\bm{y}$ might be an orientation of a galaxy, and $\bm{x}$ would be quantities such as the dark matter mass of a galaxy or the tidal field at its location. We model this distribution by employing a conditional Wasserstein generative adversarial network (GAN, \citealt{Goodfellow-GAN}).  GANs were chosen to model complex joint probability densities of all galaxies in a halo, without needing a parametric form/probability model.  

Given a generating function $g_\theta(z , \bm{x})$ with $z \sim \mathcal{N}(0, \bm{I})$, we aim to adjust the implicit distribution generated by $g_\theta$ to match our target distribution $p(\bm{y} | \bm{x})$. This can be done by minimizing the Wasserstein 1-distance $\mathcal{W}$ between these two distributions to find an optimal set of weights $\theta_{\star}$.
 By using an approximate   Wasserstein distance, we are solving the following minimax optimization problem:
\begin{equation}\label{w_dist}
	\argmin\limits_{\theta} \left( \sup_\phi \mathbb{E}_{(x, y)} \left[  d_\phi(\bm{x}, \bm{y}) -  \mathbb{E}_{z} \left[  d_\phi(g_\theta(\bm{z}, \bm{x}), \bm{y}) \right]  \right] \right)
\end{equation} 

Additionally, we must keep the Lipschitz constant bounded, to ensure that $d_\phi$ indeed parameterizes a Wasserstein distance. In   \citet{wgan}, the authors have  clipped the weights of the model to ensure the Lipschitz condition. Later \citet{wgangp} showed that the gradient constraint performs better -- thus in this work we adopt a gradient penalty.



\section{Application: Emulating Galaxy Intrinsic Alignments in Illustris-TNG simulations}
Images of distant galaxies come to us with distortions, excluding   camera and atmospheric effects. These distortions are caused by a phenomena known as weak gravitational lensing, where light  traveling from the galaxy gets deflected  due to the   presence of a massive objects (like a galaxy cluster) on the light's pathway.   
Weak lensing is measured using statistical ensembles of galaxies and their coherent  shape distortions, which are caused by the matter distribution in the Universe coherently distorting space-time. However, weak lensing measurements suffer from a number of systematics, one of which is intrinsic alignments - where galaxies are not oriented randomly in the sky in the absence of weak lensing effects, but rather tend to point towards dense regions, including those hosting other galaxies. This effect contaminates our desired weak lensing signal and can bias our measurement of dark energy. Realistic modeling of these alignments in mock galaxy catalogs is therefore paramount to validate the robustness of analysis pipelines.

\subsection{Simulated data}\label{simulation}
In this work  we are using the hydrodynamical TNG100-1 run from the IllustrisTNG simulation suite \citep[for more information, please refer to][]{ tng-bimodal,pillepich2018illustristng, Springel2017illustristng, Naiman2018illustristng, Marinacci2017illustristng,tng-publicdata}. We employ a minimum stellar mass threshold of $ \log_{10}(M_*/M_\odot) =9 $ for all galaxies, using their stellar mass from  the SUBFIND catalog.

\subsection{Graph construction}\label{graph_const} 
 
\begin{figure} \centering
\includegraphics [width=3.3 in]{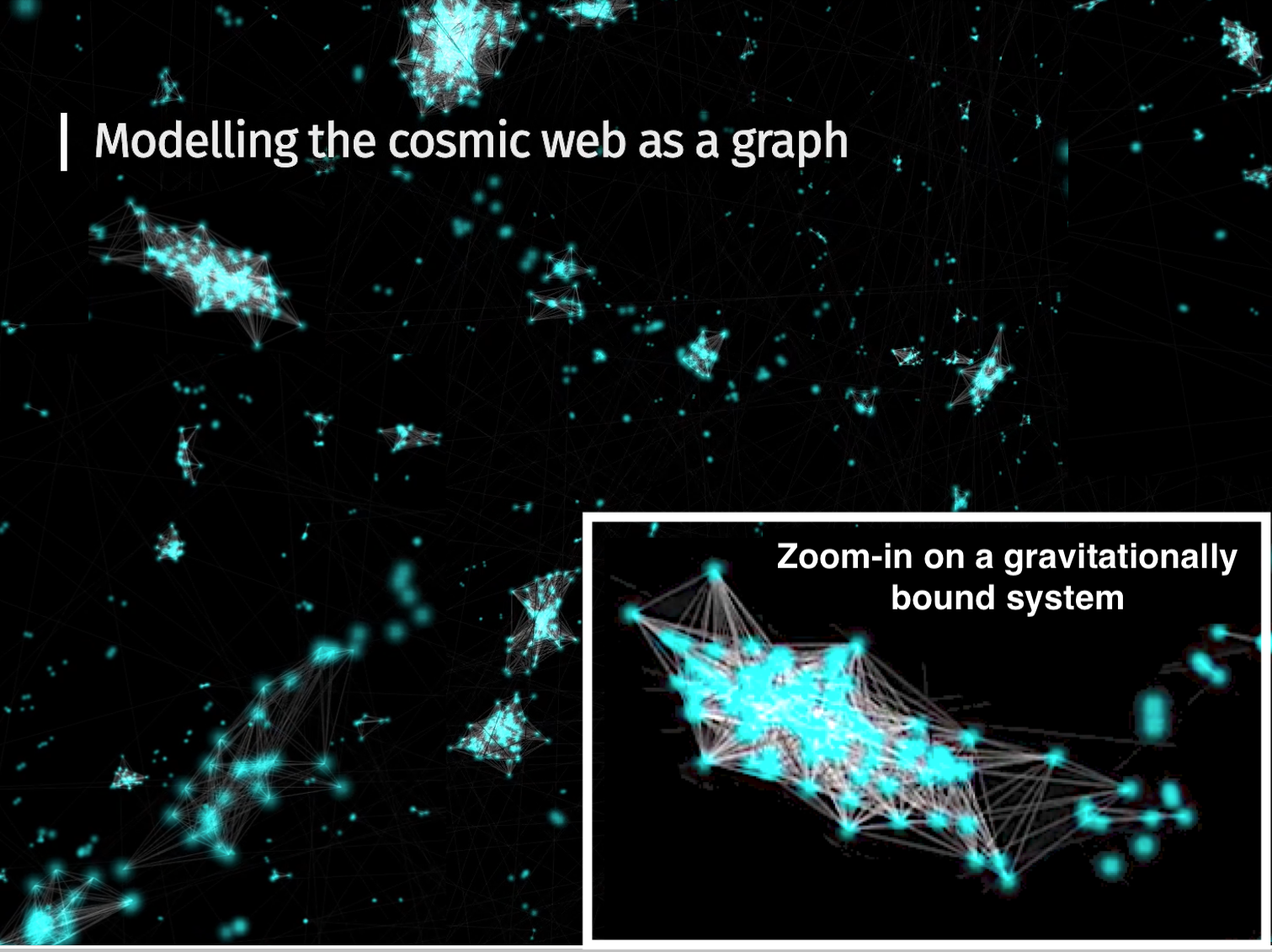}
 \caption{The cosmological simulation box modeled as a graph. Here every node represents a galaxy and the connections are made using the r-NN algorithm within each gravitationally bound system. Graph animation adapted from \citet{kim-js}.
 }\label{graph-pic}
 \end{figure}

To construct the graph for the cosmic web (i.e., for the subhalos and the galaxies), we first  grouped all subhalos and galaxies based on their parent halo.
 Given a  galaxy catalog,  an undirected graph based on
the 3D positions within the parent halo is built by placing each galaxy on a \textit{graph node}.   Then, each node has a list of features such as mass and tidal field. To build the graph connection for a given group,  the  nearest neighbors within a specified radius for a given node are connected via the \textit{undirected edges} with \textit{signals} on the graphs representing the alignments.  A snapshot of the simulation represented as \textit{graphs} is shown in Fig.~\ref{graph-pic}, where each node  represents a galaxy and the connections between the nodes are represented by grey lines.

  \begin{figure*} \centering
\includegraphics [width=6.in]{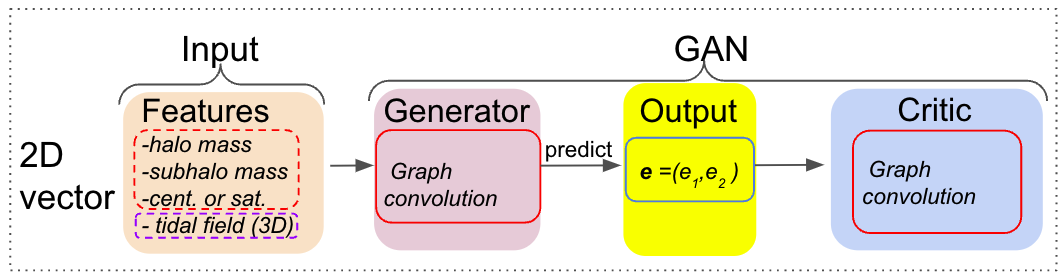}
 \caption{Architecture of the graph convolution GAN model used. Here, the input features are typical of medium to high resolution N-body (gravity only) simulations. The $\mathbf{e} = (e_1, e_2)$ is the 2D quantity that parametrizes the galaxy orientation in the sky. 
 }\label{diag}
 \end{figure*}

\subsection{Model Architecture}\label{general_arch}
 
 In Fig.~\ref{diag} we outline the architecture of our model. We have list of features (orange box) that are relevant for capturing the dependence of intrinsic alignments within a halo (dashed red box), and the tidal fields that are relevant for capturing the dependence of IA for galaxies on matter beyond their halo
 (dashed purple). These inputs are fed into the GAN-Generator (crimson box), which tries to learn the statistical distribution of the desired output labels (yellow box). At the end the input and the output from the GAN-Generator are fed into the GAN-Critic (blue box) to determine the performance of the GAN-Generator. 
 In our model, the Generator has 5 layers each with \{128,128,16,2,2\} neurons, while the Critic has 4 layers each with \{128,128,64,32\} neurons followed by a mean-pooling layer and a single output neuron.  

\subsection{Training}
 We train the model using the Adam optimizer \citep{adam} 
 with a learning rate of $10^{-3}$ and exponential decay rates of $\beta_1 = 0$ and  $\beta_2 = 0.95$. During the adversarial training we train the Generator for 5 steps and the Critic for 1 step with  a batch size of 64 (one batch is set of graphs) and a leaky ReLU activation function. As is common with GANs, our GAN models do not converge; we arbitrarily stop the training once it reached a reasonable result. 
 Our code is available at [].

\subsection{Results  }\label{results}

Throughout the section we  refer to the sample generated from the Graph-Convolutional Network-based Generative Adversarial Networks as the \textit{GAN} sample, and the sample from the TNG100 simulation as the \textit{TNG} sample.

   \begin{figure} 
\includegraphics [width=3.3in]{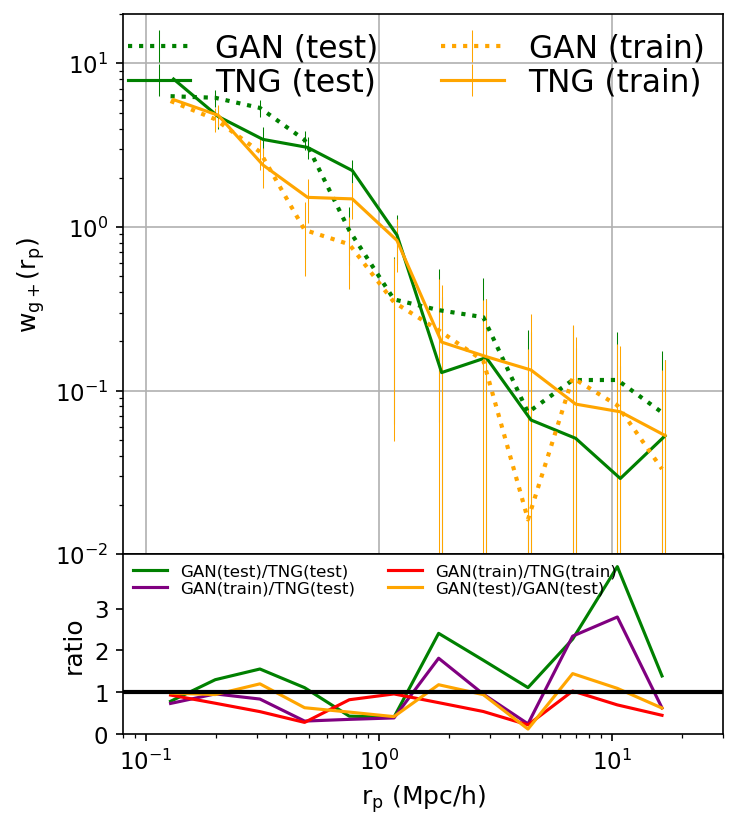}
 \caption{Projected two-point correlation functions $w_{g+}$ of galaxy positions and the projected 2D ellipticities of all galaxies, split into roughly equal-sized training and testing samples while preserving group membership. The top panel shows $w_{g+}$ as a function of projected galaxy separation $r_p$ measured using data from the TNG simulation in yellow and the data generated by the GAN in dotted green, while the bottom   panel shows the ratios among the curves as indicated by the label. All four curves are in good quantitative agreement, suggesting that the GAN is not significantly overfitting. 
 }\label{wgp_all_train_test}
 \end{figure}

For our key result, we examine $w_{g+}$, the density-shape correlation function  computed using the ellipticities (can be thought of as the 2D orientation and flattening of a galaxy when modeled as an ellipse), as shown in Fig.~\ref{wgp_all_train_test}.  The projected density-shape correlation function
$w_{g+}$ captures the correlation between overdensity and projected
intrinsic ellipticity, as is commonly used in observational studies. Positive values for $w_{g+}$ indicate that galaxies exhibit a coherent alignment towards the locations of other nearby galaxies.  
We split  our sample roughly 50/50 into training and testing samples, while preserving group membership of subhalos and galaxies.  The projected 2D correlation function, $w_{g+}$, from the GAN agrees quantitatively with the measured one from TNG simulation. 
Here, the errorbars were derived from  an analytic estimate of the covariance matrix, which includes Gaussian terms for  noise and cosmic variance (for more details see \citealt{singh-covarience,samuroff-2020}). 
Additionally, our model is also able to predict 3D  orientations and scalar quantities to a similar level of quantitative agreement.


\section{Conclusions}\label{conc}
 
 In this abstract, we have presented a novel deep generative model for scalar and vector galaxy properties. Using the TNG100 hydrodynamical simulation from the IllustrisTNG simulation suite, we have trained the model to accurately predict galaxy orientations. For a simulation box of 75 Mpc/h with 20k galaxies, the training takes about 3--4 days on a modern GPU; applying the model to a dataset of equal size is very fast (less than a minute). 
 
 Overall, the Graph Convolution based Generative Adversarial network learns and generates scalar and vector quantities that have statistical properties (distributions and alignment correlations) that agree well  with those of the original  simulation. Learning  galaxy orientations is part of a more general problem called Galaxy-Halo connection.  The problem can be stated as follows: given some properties of a dark matter halo can we predict what type of galaxy it hosts, or vice versa?  Our results represent a concrete step towards addressing this complex problem with Graph Neural Network-based Deep Generative Models.

Future work includes applying this model on a much higher volume N-body simulation with lower resolution in order to utilize  the power of deep generative models for upcoming cosmological surveys. Additionally, incorporating symmetries such as  SO(3) or E(3) and making equivariant neural networks for graphs \citep{iso-gcn,egnn} would be a useful development. 
\section*{Acknowledgements}

We thank  Ananth Tenneti, Tiziana DiMatteo, Barnabas Poczos and Rupert Croft for useful discussion that informed the direction of this work. This work was supported in part by the National Science Foundation, NSF AST-1716131 and by a grant from the Simons Foundation (Simons Investigator in Astrophysics, Award ID 620789). SS is supported by a McWilliams postdoctoral fellowship at Carnegie Mellon University. This work is supported by the NSF AI Institute: Physics of the Future, NSF PHY- 2020295.

\bibliography{main}
\bibliographystyle{icml2021}
\end{document}